\documentclass{article}
\usepackage[utf8]{inputenc}
\usepackage{arabtex}
\usepackage{textcomp}
\usepackage[numbers]{natbib}
\usepackage{mathtools}
\usepackage{float}
\title{Numerical analysis of the emission properties of terahertz photoconductive antenna by finite-difference-time-domain method}
\author{Jitao Zhang $^{\dag}$\\ ECE Department,The University of Arizona, Tucson, AZ,85721\\
$^{\dag}$ \textit{jitaozhang@email.arizona.edu}}

\date{%
    \today% thats the default I guess
    \\[2\baselineskip]% Space between date and abstract
    \normalfont\normalsize%
    \parbox{0.8\linewidth}{%
{\bfseries Abstract}: The emission properties of terahertz(THz) photoconductive antenna (PCA) have been numerically studied by three-dimensional finite-difference-time-domain method based on the full-wave model. The dependence of the THz radiation on various parameters, such as laser power, bias voltage, substrate's material, pulse duration of the laser, beam spot's size, dimension of the antenna, were comprehensively simulated and analyzed. This work, on one hand, reveals the internal relationship between the THz radiation of a PCA and the involved parameters, so that one can have a better understanding of the PCA. On the other hand, it can inspire new PCA's design that aims at improved performance, such as high radiation power, enhanced optics-to-THz conversion efficiency, and broadband spectrum.
    }
}
\usepackage{natbib}
\usepackage{graphicx}

\begin{document}

\maketitle

\section{Introduction}
Terahertz (THz) photoconductive antenna (PCA) is one of the most commonly used broadband THz source. It generates THz radiation by transient photo-excited current induced by ultrafast laser pulses\citep{auston1984}, in which several physical phenomena interact with each other in a completed manner. Even though an analytical model has been derived to try to predict the performance of a PCA\citep{duvillaret2001}, it is still highly difficult to figure out a clear picture of the influence of single involved parameter due to the mazy formula. The dependence of the emission properties of a PCA on the involved parameters can be studied by experiment and/or numerical simulation. Experimental study is a straightforward and frequently-used way to characterize the performance of a PCA. However, the preparation of the experiment is usually costly. Therefore, only part of the involved parameters can be studied when considering the availability. In addition, owing to the variety of experimental conditions (e.g. different sources of substrate materials), the experimental results from different group can not be compared fairly and sometimes even conflict with each other\citep{tani1997,reklaitis2007,liu2009,shi2009,hou2013}. Alternatively, numerical simulation is an effective and powerful route, in which the involved parameter can be studied individually and explicitly. Within the existing numerical methods\citep{fdtdel1990,fdtdsano1991,fdtdsirbu2005,fdtdkira2009,fdtdnazeri2010,drudejepsen1996,drudepiao2000,drudeduv2001,ecmholzman2000,ecmezdi2006,ecmhy2013}, the method based on the full-wave model\citep{fdtdsano1991} has least assumptions and  almost includes all of the involved parameters, which is desirable for the mentioned application.

In this work, the dependence of the emission properties of a PCA on various parameters are numerically analyzed based on the full-wave model. Using an in-house simulation tool that is developed by three-dimensional finite-difference-time-domain (FDTD) method, the influence of various parameters, such as laser power, bias voltage, substrate's material, pulse duration of the laser, beam spot's size, and dimension of the antenna, were comprehensively simulated and analyzed.

\section{Physical model and simulation method}
A typical PCA includes two parts, a semiconductor substrate and a pair of metallic electrodes deposited on the substrate, as shown in Fig.1. In order to generate THz radiation, an ultrafast
laser pulse (usually in sub-picoseconds) illuminates the region of the substrate between the electrodes, which are externally biased by a DC power supply. The photo-excited carriers (i.e. electrons and holes) inside the semiconductor is driven by the biased field to generate transient current, which then results in THz radiation in the free space with the help of the electrodes that act as an antenna.

\begin{figure}[h!]
\centering
\includegraphics[width=.8\textwidth]{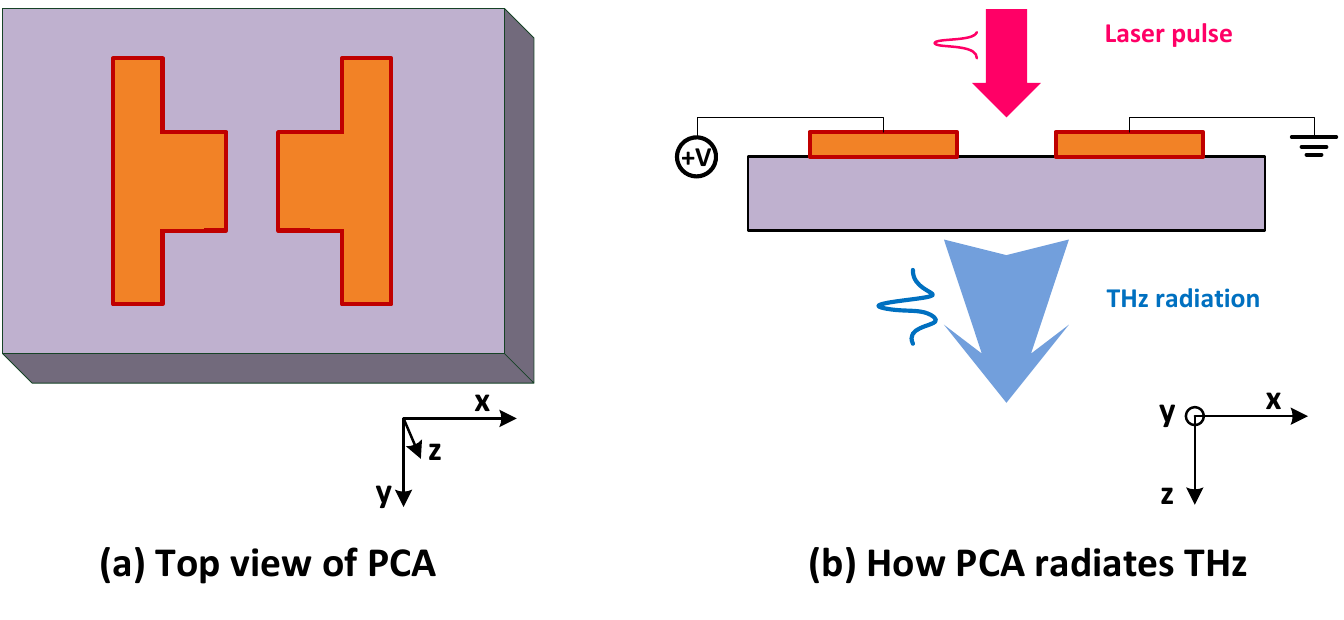}
\caption{Schematic structure of the PCA and THz radiation from the PCA. (a) A pair of dipole electrodes deposited on the surface of the semiconductor material, and (b) THz radiation from the biased PCA caused by the illumination of the laser pulse.}
\label{fig1:thz-tdssystem}
\end{figure}

The procedure of the THz radiation of a PCA can be divided into three phases: (a) build of static electric field, (b) generation of the photo-excited transient current in the near-field, and (c) THz radiation in both the near- and far-field. In the first phase, a static electric field is built inside the bulk semiconductor when the electrodes are externally biased. This static field will provide an initial field to drive the photo-excited carriers to flow towards the electrodes. The Poisson equation associated with the carrier dynamics equations  (i.e. drift-diffusion equation and continuity equation) can be applied to solve this problem, as shown in Eq.(1) $\sim$ (5):

\begin{equation}
\nabla^{2}V(\vec{r})=\frac{q}{\varepsilon}(n(\vec{r})-p(\vec{r})-N_{D}+N_{A})
\end{equation}

\begin{equation}
\nabla\cdot J_{n}(\vec{r})=qR
\end{equation}

\begin{equation}
\nabla\cdot J_{p}(\vec{r})=-qR
\end{equation}

\begin{equation}
J_{n}(\vec{r})=q\mu_{n}n(\vec{r})(-\nabla\cdotp V(\vec{r}))+qD_{n}\nabla n(\vec{r})
\end{equation}

\begin{equation}
J_{p}(\vec{r})=q\mu_{p}p(\vec{r})(-\nabla\cdotp V(\vec{r}))-qD_{p}\nabla p(\vec{r}),
\end{equation}

\noindent where \textit{V} is the voltage distribution inside the semiconductor, \textit{q} is elementary charge, $\varepsilon $ is permittivity of the semiconductor, \textit{n} and \textit{p} are density of electrons and holes, respectively, $ N_{D}-N_{A} $ represents the concentration of impurities, $J_{n}$ and $J_{p}$ are current density of electrons and holes, respectively, $R$ is the recombination rate of the carriers, $\mu_{n}$ and $\mu_{p}$ are mobilities of the carriers, and $D_{n}$ and $D_{p}$ are diffusion coefficients, which are related to the mobilities by Einstein relationship
\begin{equation}
\frac{D_{n}}{\mu_{n}}=\frac{D_{p}}{\mu_{p}}=\frac{K_{B}T}{q}.
\end{equation}
The $\vec{r}$ indicates that the corresponding parameters are vectors. By solving the above equations, we can obtain the steady solution of the electric field ($E_{DC}$), carrier densities  ($n_{DC}$ and $p_{DC}$)  and current density ($J_{n_{DC}}$ and $J_{p_{DC}}$) inside the semiconductor for the first phase. In the second phase, a transient current is generated when a laser pulse illuminate the PCA's gap according to the carrier dynamics model. In the third phase, the transient current will result in THz radiation through the electrodes, which can be predicted by Maxwell's equation. The coupling between phase (b) and phase (c) is realized by using the transient current as driving source of the antenna to update the electromagnetic field. The physical model used here is summarized in eq.(7) $\sim$ (14):

\begin{equation}
\nabla\times E(\vec{r})=-\mu \frac{\partial H(\vec{r}) }{\partial t}
\end{equation}

\begin{equation}
\nabla\times H(\vec{r})=\varepsilon \frac{\partial E(\vec{r}) }{\partial t}+J_{n,pho}(\vec{r})+J_{p,pho}(\vec{r})
\end{equation}

\begin{equation}
q\frac{\partial n(\vec{r})}{\partial t} = \nabla\cdot J_{n}(\vec{r})+q(G-R)
\end{equation}

\begin{equation}
q\frac{\partial p(\vec{r})}{\partial t} = - \nabla\cdot J_{p}(\vec{r})+q(G-R)
\end{equation}

\begin{equation}
J_{n}(\vec{r})=q\mu_{n}n(E_{DC}(\vec{r})+E(\vec{r}))+qD_{n}\nabla n(\vec{r})
\end{equation}

\begin{equation}
J_{p}(\vec{r})=q\mu_{p}p(E_{DC}(\vec{r})+E(\vec{r}))-qD_{p}\nabla p(\vec{r})
\end{equation}

\begin{equation}
J_{n,pho}(\vec{r})=J_{n}(\vec{r})-J_{n_{DC}}(\vec{r})
\end{equation}

\begin{equation}
J_{p,pho}(\vec{r})=J_{p}(\vec{r})-J_{p_{DC}}(\vec{r}),
\end{equation}

\noindent where $E$ and $H$ are radiated electric and magnetic field, respectively,$\mu$ is permeability, $J_{n,pho}$ and $J_{p,pho}$ specifically represent the photo-excited current density, and $G$ is the generation rate of the photo-excited carriers. Other symbols have the same meanings as above. The continuity equation shown in Eq.(9) and (10) describes the carrier dynamics, and drift-diffusion equation shown in Eq.(11) and (12) describes the corresponding transient current. By using photo-excited current (Eq.(13) and (14)) as driving source of an antenna to update the Maxwell's equation (Eq.(7) and (8)), the THz radiation can be precisely predicated, both in the near-field and far-field. The numerical implementation of the above model by FDTD method can be found in our previous work\citep{jitao2014}.

A coplanar stripline PCA with a total size of $50\mu m\times 50\mu m\times 2.2\mu m$ was used for numerical simulation. The detailed dimension is shown in Fig.\ref{fig:dimension_both}(a). A laser beam (red circle) with diameter of $20\mu m$ is placed near the anode. The low-temperature grown gallium arsenide(LT-GaAs) was used as substrate's material. The permittivity of GaAs is $\varepsilon =12.9$, and the absorption coefficient is $1\times 10 ^{4} cm^{-1}$. The intrinsic concentration is $2.1\times 10^{6} cm^{-3}$ . The carrier lifetimes of the electron and hole are $0.1 ps$ and $0.4 ps$, respectively. The mobilities of the electron and hole are $200 cm^{2}/V\cdot s$ and $40 cm^{2}/V\cdot s$, respectively. The laser has a wavelength of $800 nm$, pulse duration of $80 fs$ and beam size of $20 \mu m$. The far-field distance is $200 mm$ away from the center of the PCA right below the substrate.

\section{Simulation results of parameter studies}
\begin{figure}[H]
\centering
\includegraphics[width=.7\textwidth]{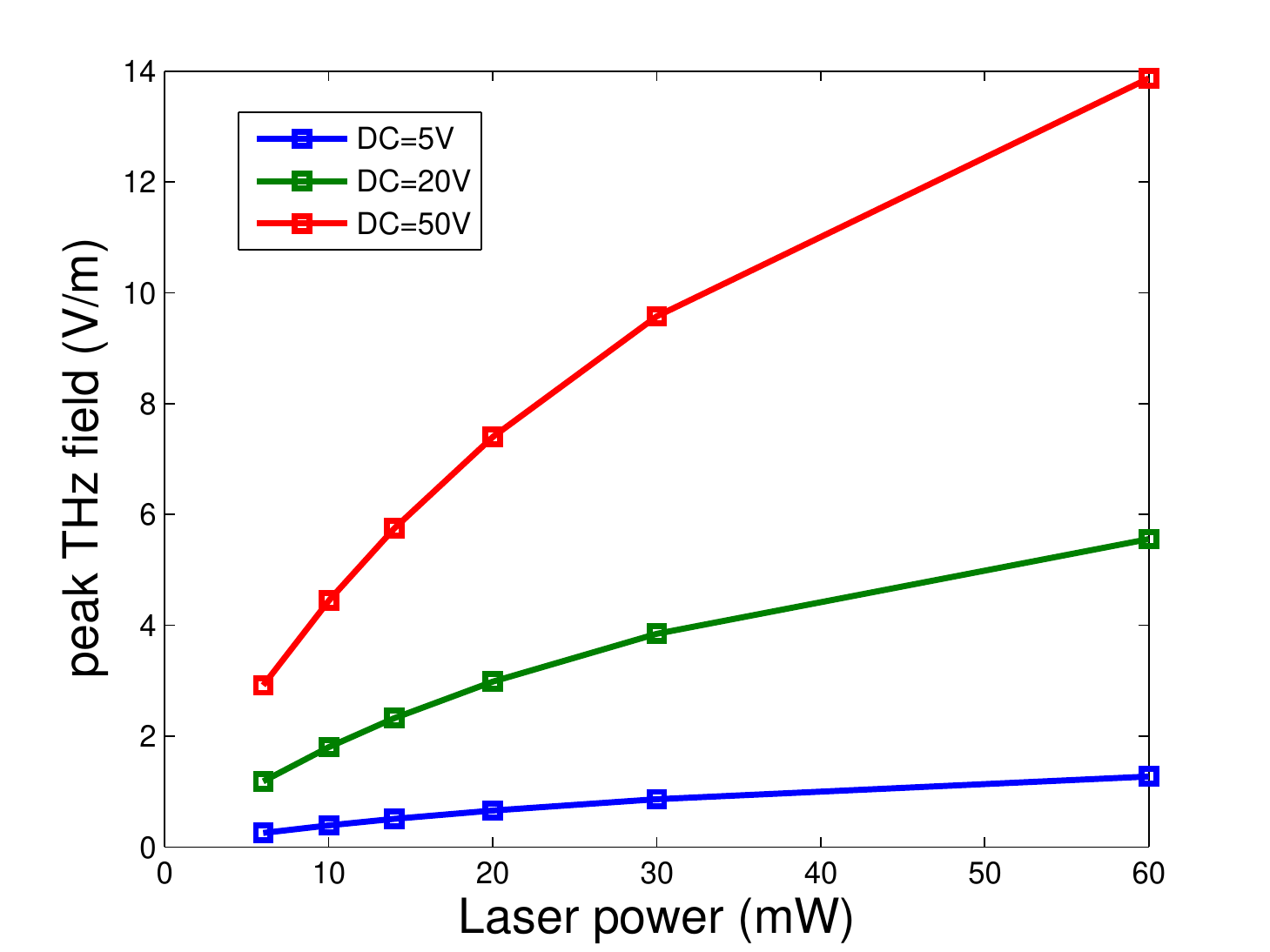}
\caption{Dependence of the THz field on the laser power for LT-GaAs substrate. The squares represent the simulation data, and the solid lines are guide of eye.}
\label{fig:lt-gaas-laser}
\end{figure}

\begin{figure}[H]
\centering
\includegraphics[width=.7\textwidth]{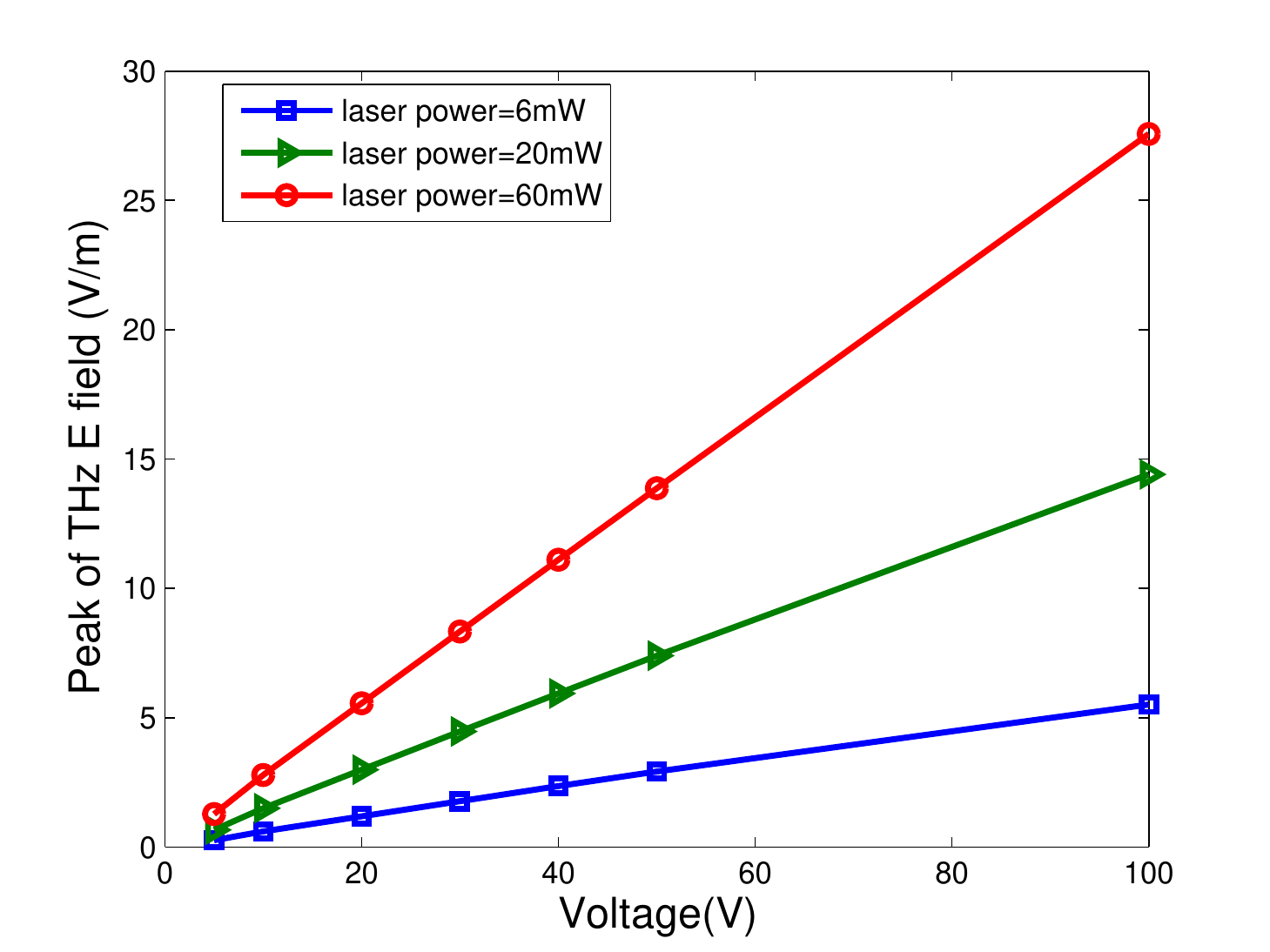}
\caption{Dependence of the THz field on the bias voltage for LT-GaAs substrate. The squares,triangles, and circles represent the simulation data, and the solid lines are guide of eye.}
\label{fig:lt-gaas-voltage}
\end{figure}

We first studied the dependence of the THz radiation in far-field on laser power and bias voltage. For laser power study, the laser power was varied from $6 mW$ to $60 mW$ at three bias voltages($5V$, $20V$, and $50V$). For bias voltage study, the bias voltage was varied from $5V$ to $100V$ at three laser powers ($6mW$,$20mW$, and $60mW$ ). The peak of the THz field with regard to the laser power and bias voltage are shown in Fig.\ref{fig:lt-gaas-laser} and Fig.\ref{fig:lt-gaas-voltage}. The saturation effect of the THz field against the laser power can be clearly found in Fig.\ref{fig:lt-gaas-laser}. In addition, A linear relationship between the THz field and the bias voltage can be found in Fig.\ref{fig:lt-gaas-voltage}. Both are consistent to the prediction of the scaling rule\citep{darrow1992,ben1993}.

\begin{figure}[H]
\centering
\includegraphics[width=.7\textwidth]{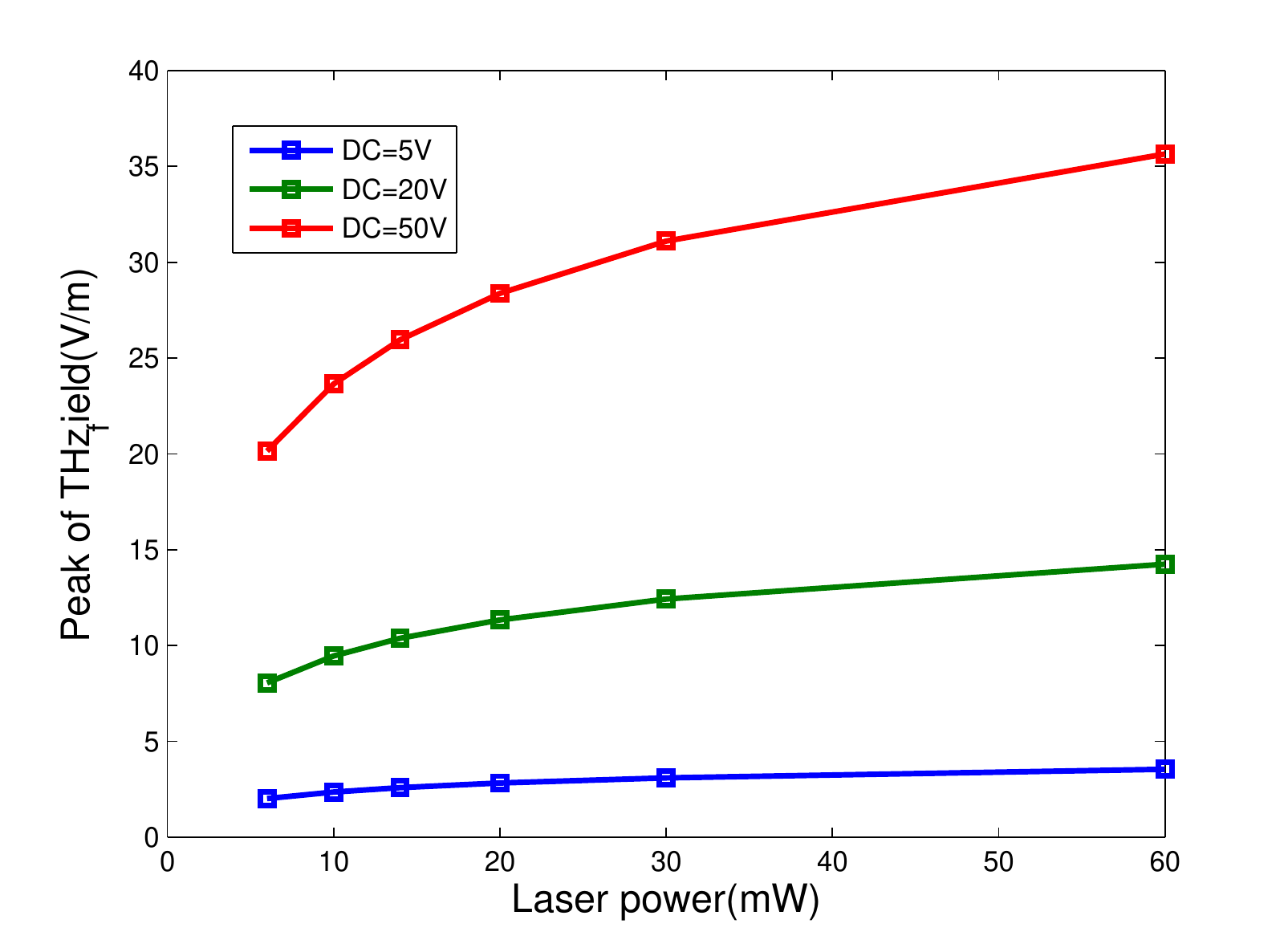}
\caption{Dependence of the THz field on the laser power for SI-GaAs substrate. The squares represent the simulation data, and the solid lines are guide of eye.}
\label{fig:si-gaas-laser}
\end{figure}

\begin{figure}[H]
\centering
\includegraphics[width=.7\textwidth]{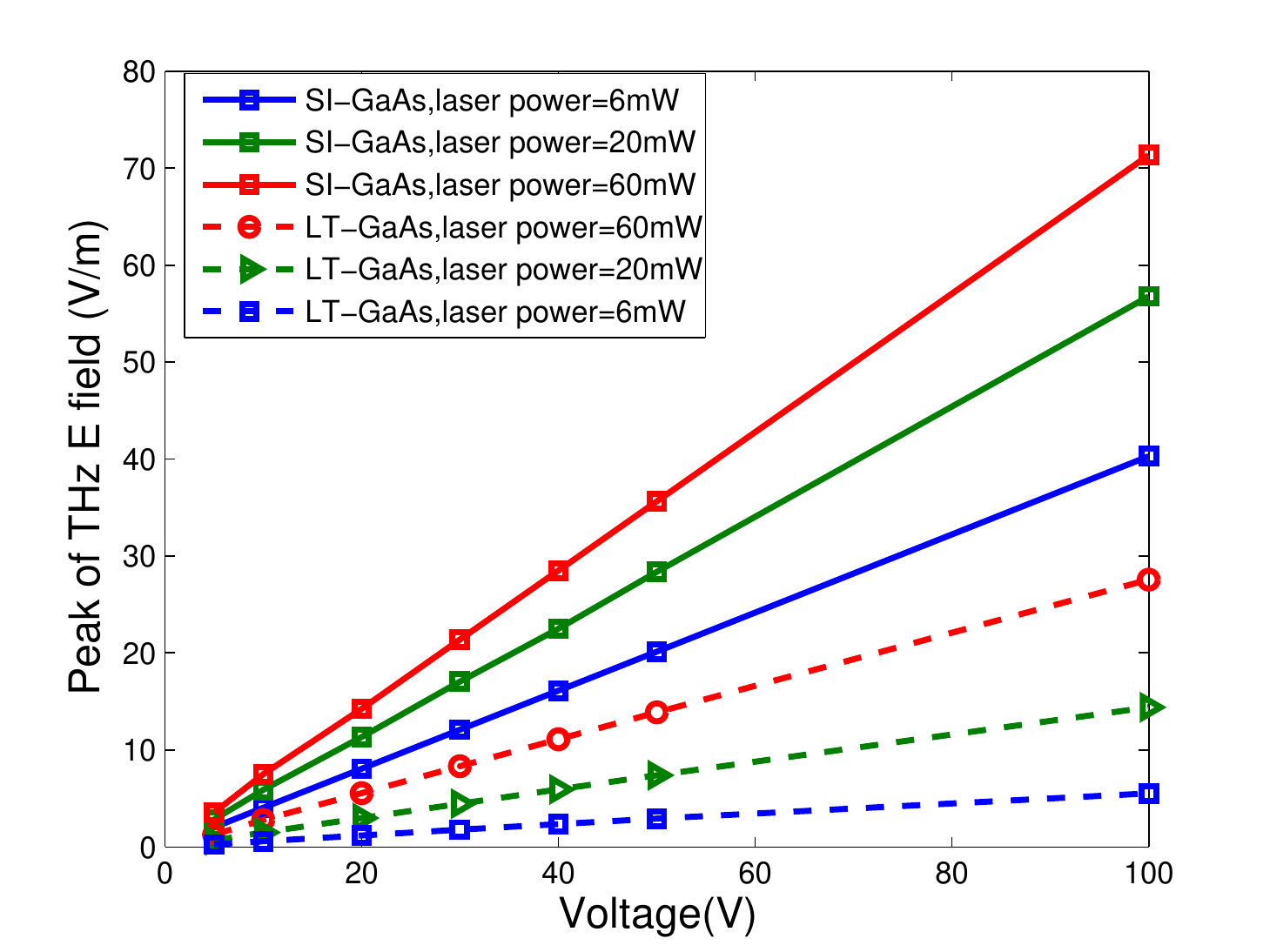}
\caption{Dependence of the THz field on the bias voltage for LT-GaAs and SI-GaAs substrates. The squares,triangles, and circles represent the simulation data, and the solid and dash lines are guide of eye.}
\label{fig:si-gaas-voltage}
\end{figure}

In order to evaluate the effect of substrate's material, Semi-Insulating GaAs (SI-GaAs) was used for similar study as shown above. The mobility of the electron and hole in SI-GaAs are $5000 cm^{2}/V\cdot s$ and $200 cm^{2}/V\cdot s$, respectively, and the carrier lifetime of electron and hole are $10ps$ and $40ps$, respectively. The results are shown in Fig.\ref{fig:si-gaas-laser} and Fig.\ref{fig:si-gaas-voltage}. The features of SI-GaAs based PCA in Fig.\ref{fig:si-gaas-laser} and \ref{fig:si-gaas-voltage} are similar to that of the LT-GaAs based PCA, except the THz radiation is stronger at the same laser power and bias voltage. The data of LT-GaAs based PCA was also drawn in Fig.\ref{fig:si-gaas-voltage} as a comparison. 

It was found that the large mobility of SI-GaAs substrate accounted for the enhanced THz field. This can be further understood based on Eq.(11), in which the photo-excited current is approximately proportional to the carrier's mobility in case the diffusion current is small enough (This is usually the case). Similar enhancement has also been found in experiment\citep{tani1997}.

\begin{figure}[H]
\centering
\includegraphics[width=0.7\textwidth]{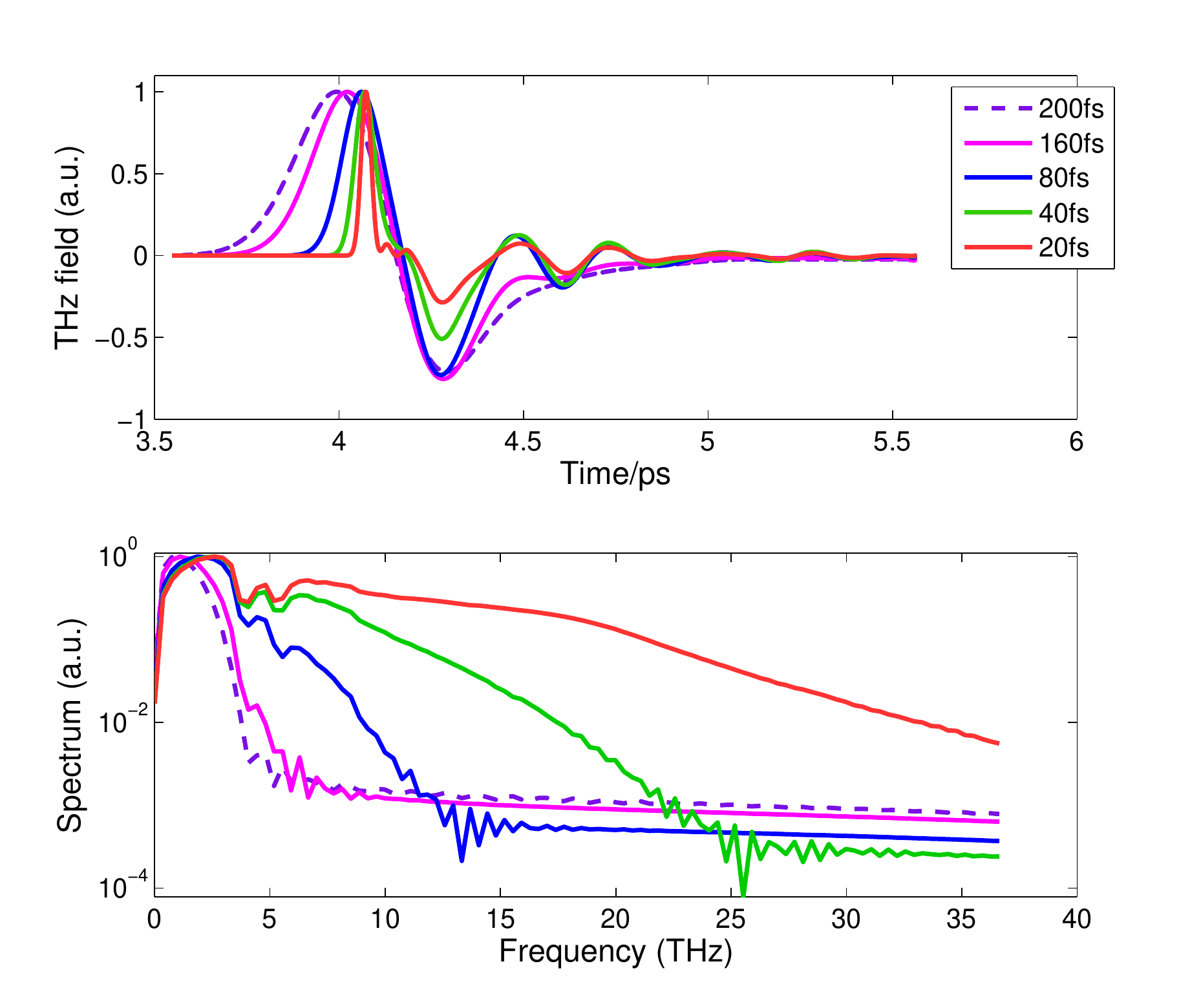}
\caption{Dependence of the THz field on the pulse duration of the laser. (Upper) time domain THz field,(Lower) corresponding spectrum by Fast-Fourier-Transformation.}
\label{fig:pulseduration}
\end{figure}

The THz radiation of a PCA is a result of the transient current, which is excited by the incoming ultra-fast laser pulse. In order words, the pulse duration of the laser will affect the temporal shape of the transient current, and thus, affect the bandwidth of the THz radiation. For this study, the pulse duration of the incoming laser was varied from $20fs$ to $200fs$ at laser power of $20mW$ and bias voltage of $60V$, and the corresponding THz radiations of the LT-GaAs based PCA in far-field were calculated and compared, as shown in Fig.\ref{fig:pulseduration}. 

It indicates that the bandwidth of the THz radiation depends on the pulse duration of the laser monotonously. For example, when excited by $20fs$ pulse, the PCA can radiate THz field beyond $30THz$. As the pulse duration was expanded to $200fs$, the bandwidth of the PCA decreased to less than $5THz$. In all cases, even though the bandwidth varies dramatically, the peak frequencies always lie in $3THz$, which is determined by the carrier lifetime of the LT-GaAs.

\begin{figure}[H]
\centering
\includegraphics[width=.7\textwidth]{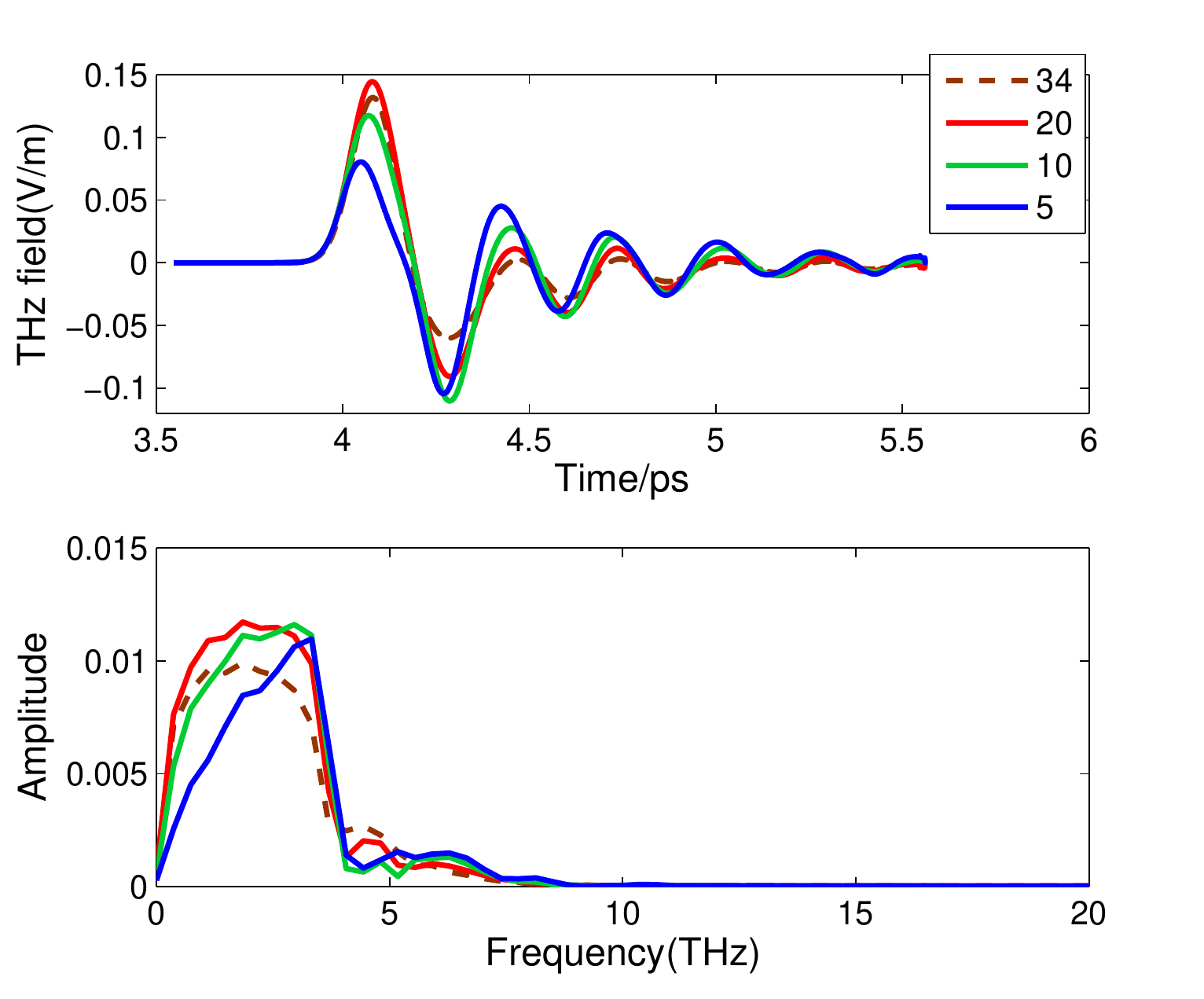}
\caption{Effect of the spot's size of the laser beam. The labels indicate the spot's size in micrometer. (Upper) time domain THz field,(Lower) corresponding spectrum by Fast-Fourier-Transformation.}
\label{fig:beamsize}
\end{figure}

The spot's size of the laser beam determines an illuminated region, in which the photo-excited carrier will be generated. It is interesting to figure out the influence of the spot's size. For this purpose, three spot's sizes ($5\mu m$,$10\mu m$, $20\mu m$, and $34\mu m$) were used for simulation at the same laser power($2.6mW$) and bias voltage ($5V$). The laser beams always located at the vicinity of the anode. The result is shown in Fig.\ref{fig:beamsize}. It indicates that larger beam spot can result in better THz radiation, in both time-domain and frequency-domain. This is probable caused by the screening effect. For the same input laser power, smaller spot has higher optical intensity, and thus, has higher current density. This results in more intense localized field, which will screen the bias field so that the THz radiation is decreased. On the contrary, the screening effect will be weaker for larger beam spot, so that the THz radiation is more effective. However, when the beam spot is as large as overlapping the whole gap, the THz radiation will again become weaker. In this case, the average transport path to the electrode is so large that part of the photo-excited carriers will be recombined before arriving at the electrode.

\begin{figure}[H]
\centering
\includegraphics[width=.8\textwidth]{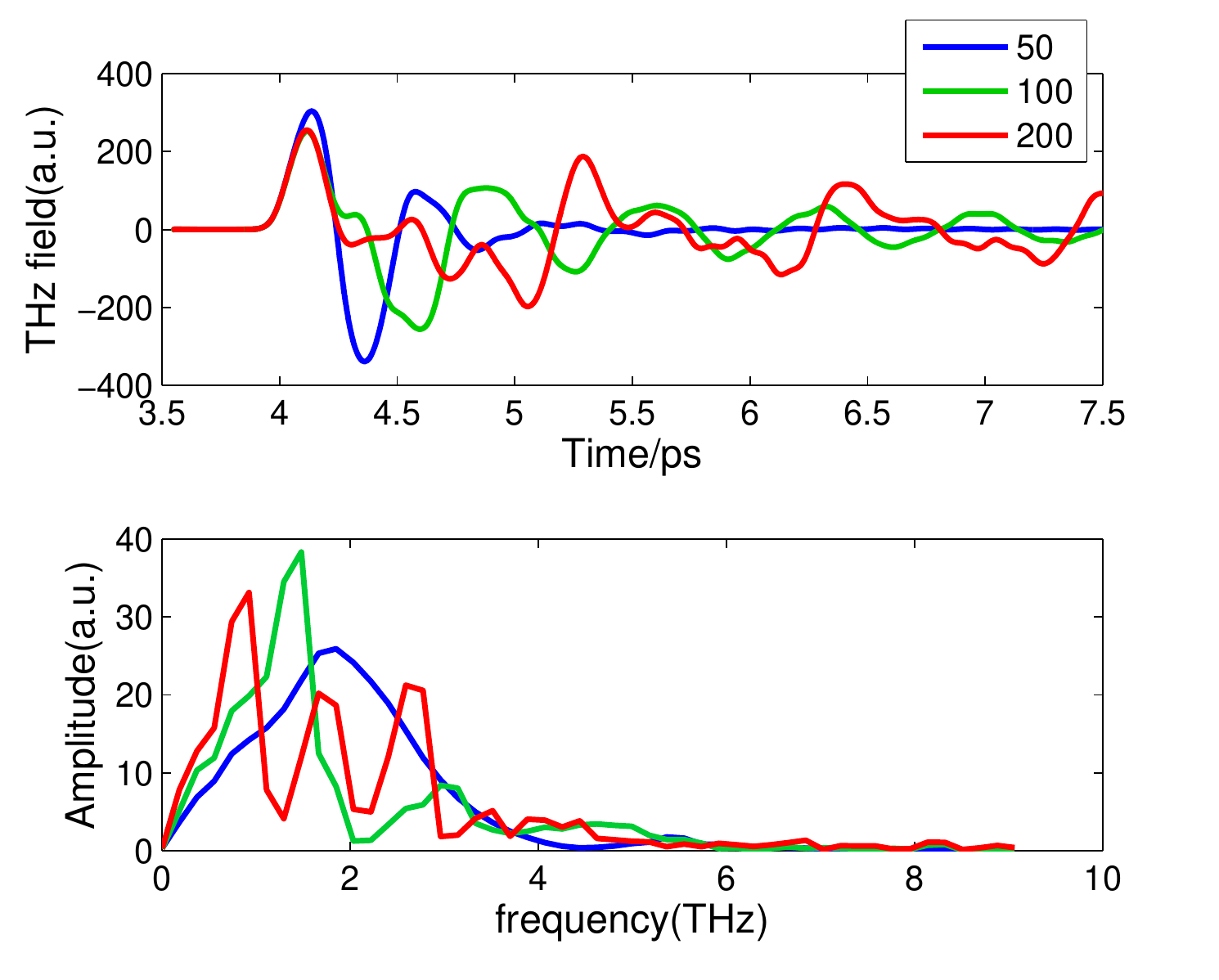}
\caption{Dependence of the THz field on the length of transmission line of a dipole PCA. The labels indicate the length of the transmission line in micrometer.}
\label{fig:line_length}
\end{figure}

The free-space THz radiation of a PCA is mainly due to the electrode's structure, which works as an antenna. In other words, the electrode's structure will affect the emission property of a PCA in some way. We studied this effect by varying the length of the transmission line of a dipole PCA, whose dimension is shown in Fig.\ref{fig:dimension_both}(b). The dipole has a size of $20\mu m\times 40\mu m$, and the transmission line has a width of $5\mu m$. The laser power and bias voltage were fixed at $2.6mW$ and $5V$, respectively. The length of the transmission line were set as $50\mu m$,$100\mu m$, and $200\mu m$, and the corresponding THz radiation in far-field were simulated and compared, as shown in Fig.\ref{fig:line_length}. The time-domain THz field of the PCA with longer transmission line shows reflected signal after the primary pulse, which is corresponded to the length of the transmission line. In addition, the spectra show \textit{red-shift} against the length of the transmission line.

\section{Conclusion}

The emission properties of THz PCA have been analyzed by numerical simulation. The dependence of the THz radiation in far-field on various parameters, such as laser power, bias voltage, substrate's material, beam spot's size, pulse duration of the laser, and the dimension of the antenna, were studied comprehensively. In order to enhance THz radiation, the semiconductor having larger mobility and higher breakdown tolerance can be used as PCA's substrate. In addition, larger beam spot that comparable to the gap size can generate THz radiation more effectively, and short pulse duration of the laser can generate broadband THz field. Moreover, the length of the transmission line will affect the bandwidth of the THz field, which should be considered during the design of a PCA.

\section*{Acknowledgements}
This work can not be done without the contribution of the following colleagues. They are Mingguang Tuo, Min Liang and Hao Xin. They will be listed as co-authors when we consider to publish this work in a journal.

\appendix
\section{Dimension of the PCAs }

\begin{figure}[H]
\centering
\includegraphics[width=.7\textwidth]{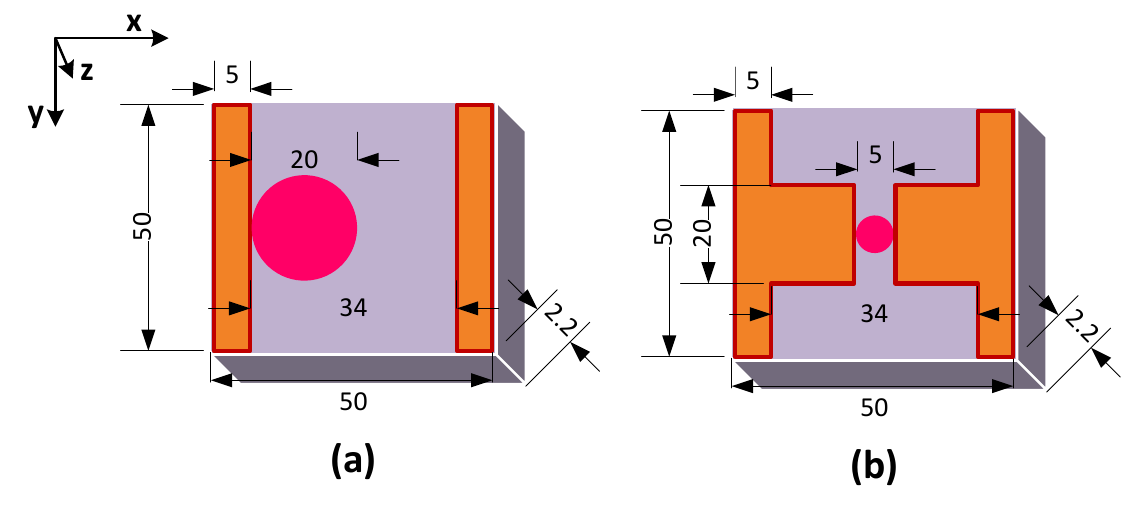}
\caption{Dimension of the (a)coplanar stripline PCA and (b)dipole PCA, all dimensions are in micrometer. The solid circle indicates the location of the laser beam.}
\label{fig:dimension_both}
\end{figure}

\bibliographystyle{plain}
\bibliography{references}
\end{document}